\documentclass[prl,showpacs,notitlepage,twocolumn]{revtex4}
\usepackage{amsmath}
\usepackage{amssymb}
\usepackage{bm}
\usepackage{epsfig}
\usepackage{graphicx}
\usepackage{color}
\usepackage{natbib}
\usepackage{hyperref}
\usepackage{multirow} 
\newcommand {\e}{{\rm e}}
\renewcommand {\i}{{\rm i}}
\newcommand{\nix}[1]{}
\begin{document}
  \title{Anomalous Suppression of Valley Splittings in Lead Salt Nanocrystals without
Inversion Center }

\author{A.N. Poddubny,${}^1$ M.O. Nestoklon,${}^1$ S.V. Goupalov${}^{1,2}$}
 
 \affiliation{${}^1$Ioffe Physical-Technical Institute,
 26 Polytekhnicheskaya St., 194021 St.~Petersburg, Russia\\
${}^2$ Department of Physics, Jackson State University, Jackson, Mississippi 39217, USA}

\begin{abstract}
Atomistic $sp^3d^5s^*$ tight-binding theory of PbSe and PbS
nanocrystals is developed. It is demonstrated, that the valley
splittings of confined electrons and holes strongly and peculiarly
depend on the geometry of a nanocrystal.
When the nanocrystal lacks a microscopic center of inversion and has $T_d$
symmetry, the splitting is strongly suppressed as compared to the more symmetric nanocrystals with $O_h$ symmetry, having an inversion center.
\end{abstract}
\maketitle


Interest in almost spherical nanocrystals (NCs) made of lead chalcogenides (PbSe, PbS) has recently
exploded due to their utility in fundamental studies of quantum confinement effects
and enabling potential for technological applications in photovoltaics~\cite{ellingson,schaller,inspite,beard}.
These applications impose requirements of reproducibility
of the devices, ability to control their properties and necessity to understand mechanisms
behind physical processes underlying their work.
It has been noted~\cite{Delerue2004b,zunger2006} that lead salt NCs
are very peculiar compared to quantum dots (QDs) of III-V and II-VI compound semiconductors because
lead chalcogenides have band extrema in the four inequivalent $L$-points of the Brillouin
zone and such
effects as confinement-induced valley-mixing and effective mass anisotropy should be considered to fully
account for the properties of lead salt NCs. 
 It also was suggested that optical properties of lead salt NC QDs can
be sensitive to a particular arrangement of atoms within the QD,
as the latter determines the overall symmetry of the
structure~\cite{goupalov2009,nootz2010}.

In this work we study how all these effects influence the valley-orbit and spin-orbit splittings
of one-particle energy levels of electrons and holes confined in lead salt NCs using the
atomistic tight-binding (TB) approach. We found that
these splittings are very sensitive to the particular arrangement of atoms within the almost
spherical NC. In particular, we considered NCs of almost
spherical shape
centered on an anion or cation atom, serving as a center of inversion,
along with NCs having
no inversion symmetry. We found that in NCs without a center of inversion the valley-orbit and spin-orbit
splittings of electron energy levels are strongly suppressed. This effect is quite unusual because typically a higher symmetry
of a physical system implies a higher degeneracy of its energy levels, while in our case the suppression
of the splittings occurs in NCs having lower symmetry. Nevertheless, we were able to explain this puzzling
behavior using mathematical apparatus of the group theory.



Lead chalcogenides (PbSe, PbS) are semiconductor compounds
with a rocksalt crystal lattice and a narrow and direct band gap~\cite{Kang1997}.
The extrema of both the conduction and valence bands are
located at the four $L$-points of the Brillouin zone:
\begin{equation}\label{eq:kL}
{\bm k}_{1,2}=\tfrac{\pi}{a}\left(1,\pm 1,\pm 1\right), \,
{\bm k}_{3,4}=\tfrac{\pi}{a}\left(-1,\pm 1,\mp 1 \right), \,
\end{equation}
where $a$ is the lattice constant.
The empirical TB method is an efficient tool to model electronic
properties of large-scale nanostructures~\cite{Delerue2004}.
Success of the TB parametrization depends on the choice of basis
functions and on the accuracy of the fit of the bulk band structure.
The simplest  TB parametrizations of
lead chalcogenides are based on the basis set of the three $p$
orbitals playing major role in the formation of
the valence and conduction band states~\cite{Mitchell1966,Volkov1983}.
More quantitatively accurate models include also
$s^*$ and $d$ orbitals~\cite{Lent1986,Lachhab2002,Valdivia1995,Delerue2004b}.
However, no attempts (with the only exception of Ref.~\cite{Volkov1983})
have been made to fit the actual effective masses of the electrons
and holes near the $L$-points.
On the other hand, the second-nearest neighbors $p^3$
model of Ref.~\cite{Volkov1983}
fails to reproduce the bulk dispersion for wavevectors far from the $L$
points~\cite{zunger2006}.
Consequently, even the most advanced existing TB parametrizations of
lead chalcogenides\cite{Delerue2004b} are not
suitable\cite{klimeck2011} for
an adequate description of the NCs.

We have performed an independent atomistic $sp^3d^5s^*$
TB parametrization
of the electron energy dispersion in bulk PbSe and PbS by fitting the spectra
calculated by the state-of-the art GW technique of Ref.~\cite{Cardona2010}.
The goal values for the carrier effective masses near the
$L$-points
were set to the experimental values~\cite{preier1979}:
$
 m_{c,l}^{exp}=0.070 \, m_0,\: m_{c,t}^{exp}=0.040 \, m_0,\:$
 $m_{h,l}^{exp}=0.068 \, m_0,\: m_{h,t}^{exp}=0.034 \, m_0
$
for PbSe and
$
m_{c,l}^{exp}=0.105 \, m_0,\: m_{c,t}^{exp}=0.080 \, m_0,$ $m_{h,l}^{exp}=0.105 \, m_0,\: m_{h,t}^{exp}=0.075 \, m_0$
for PbS ($m_0$ is the free electron mass),
as even the modern
{\it ab initio} approach\cite{Cardona2010}
does not satisfactory reproduce the effective masses.

\begin{table}[b!]
 \caption{TB parameters for PbSe and PbS.
The transfer integrals are meausured in eV and given in the
Slater-Koster notations~\cite{SlaterKoster}.
The spin-orbit splittings are defined according to Ref.~\cite{chadi1977}.
}\label{table}
\begin{center}
\begin{tabular}{c|cc}
  & PbS& PbSe\\\hline
  $a_0$, \AA& $5.900$& $6.100$\\
  $E_{sa}$& $-10.596$& $-10.722$\\
  $E_{sc}$& $-5.444$& $-6.196$\\
  $E_{pa}$& $-1.797$& $-1.463$\\
  $E_{pc}$& $4.819$& $4.279$\\
  $E_{da}$& $7.468$& $7.984$\\
  $E_{dc}$& $20.900$& $26.114$\\
  $E_{s^*a}$& $17.878$& $15.117$\\
  $E_{s^*c}$& $25.807$& $28.244$\\
  $ss\sigma$& $-0.567$& $-0.292$\\
  $s^*s^*\sigma$& $-2.478$& $-1.346$\\
  $s_cs^*_a\sigma$& $-1.535$& $-0.654$\\
  $s_as^*_c\sigma$& $-0.693$& $-1.743$\\
$s_ap_c\sigma$& $1.623$& $1.611$\\
  $s_cp_a\sigma$& $1.371$& $1.291$\\
$\Delta_a$& $0.096$& $0.420$\\
\end{tabular}
\begin{tabular}{c|cc}
  & PbS& PbSe\\\hline
  $s^*_ap_c\sigma$& $2.606$& $2.258$\\
  $s^*_cp_a\sigma$& $2.177$& $1.731$\\
  $s_ad_c\sigma$& $-1.852$& $-1.917$\\
  $s_cd_a\sigma$& $-1.399$& $-1.256$\\
  $s^*_ad_c\sigma$& $0.040$& $0.146$\\
  $s^*_cd_a\sigma$& $-0.792$& $-0.271$\\
  $pp\sigma$& $2.223$& $2.159$\\
  $pp\pi$& $-0.468$& $-0.463$\\
  $p_ad_c\sigma$& $-1.200$& $-1.272$\\
  $p_c d_a \sigma$& $-1.219$& $-1.332$\\
  $p_ad_c\pi$& $0.442$& $0.912$\\
  $p_c d_a \pi$& $0.983$& $0.966$\\
  $dd\sigma$& $0.778$& $0.244$\\
  $dd\pi$& $1.202$& $1.826$\\
  $dd\delta$& $-1.305$& $-1.235$\\
   $\Delta_c$& $2.380$& $2.380$\\
\end{tabular}
\end{center}
\end{table}

The TB parameters we obtained
are listed in Table~\ref{table}.
The resulting effective masses
$
 m_{c,l}=0.068 \, m_0,\: m_{c,t}=0.041 \, m_0,$ $
 m_{h,l}=0.069 \, m_0,\: m_{h,t}=0.039 \, m_0
$
for PbSe and
$
 m_{c,l}=0.098 \, m_0,\: m_{c,t}=0.079 \, m_0,$\:$
 m_{h,l}=0.104 \, m_0,\: m_{h,t}=0.074 \, m_0
$
for PbS are quite close to the experimental values.
The spin-orbit coupling constants of $p$ orbitals at Pb, Se, and S
were not changed during the fitting procedure and were taken from
Refs.~\cite{Herman1963}
and~\cite{HermanSkillman1963} for Pb and
for the anions, respectively.

\begin{figure}[t]
 \begin{center}
  \includegraphics[width=0.45\textwidth]{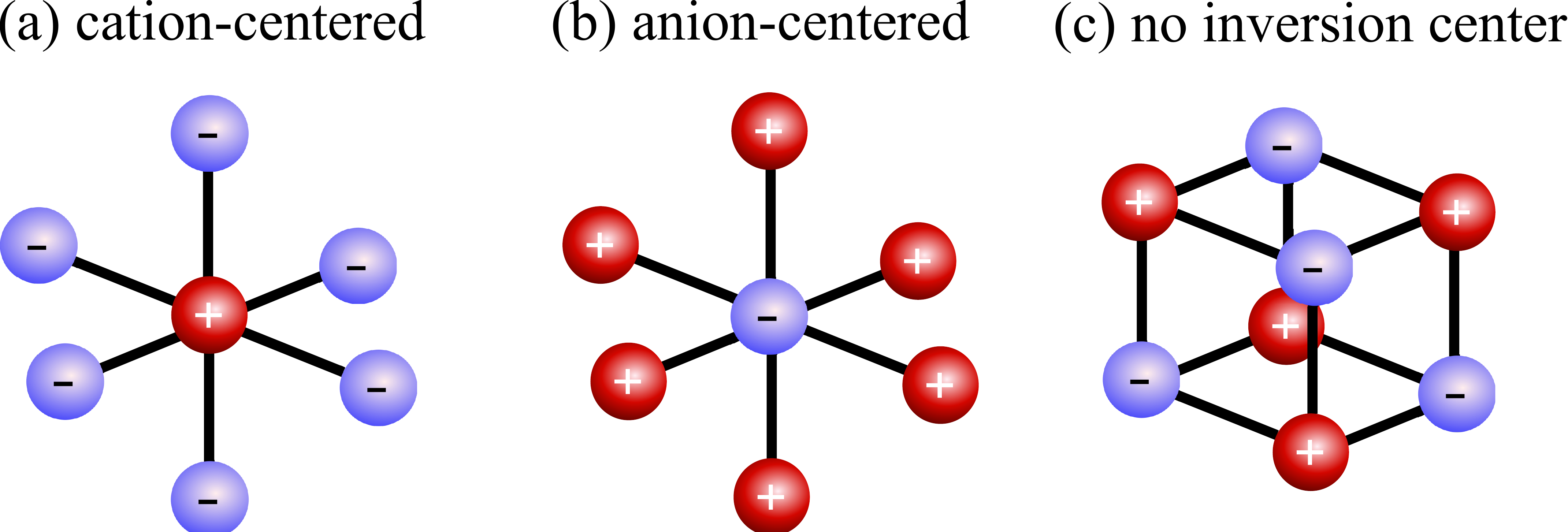}
 \end{center}
\caption{ Central parts of the three types of
nanocrystals.}\label{fig:structure}
\end{figure}

%
In our study we will consider the three types of NC geometries
illustrated
in~Fig.~\ref{fig:structure}.
\begin{figure}[ht!]
\begin{center}
 \includegraphics[width=\linewidth]{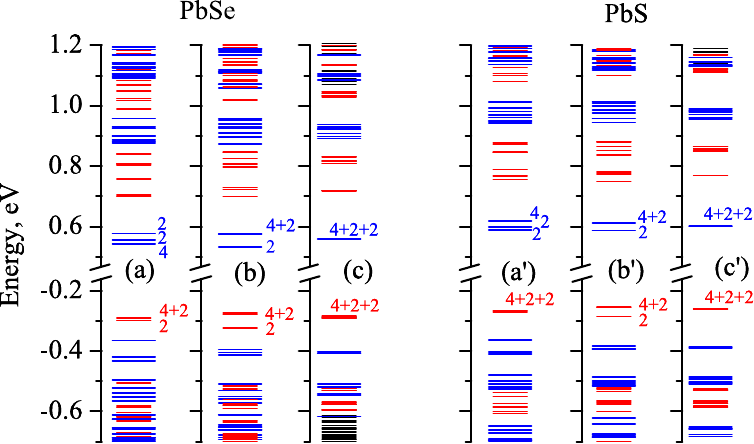}
\end{center}
 \caption{(Color online)
Energy levels in  PbSe (a,b,c)  and PbS (a$'$,b$'$,c$'$) NCs with
diameter $D\approx 4.9$~nm  and  $D\approx 4.6$~nm, respectively.
Panels (a--c) and (a$'$--c$'$) correspond to cation-centered NCs,
anion-centered NCs and NCs without an inversion center,
respectively (see Fig.~\ref{fig:structure}).
States of the odd and even parity are marked by the
long thick blue and short thick red lines, respectively.
The thin black lines correspond to the states without a certain parity.
The degeneracies of the lowest electron and hole confined states
are indicated near each line.
}\label{fig:Dos}
\end{figure}
For the structures shown in~Figs.~\ref{fig:structure}a(b)
 the center of the
spherical NC is on a cation (anion) atom while for the structure
in~Fig.~\ref{fig:structure}c the center of the sphere lies halfway
between a cation and an anion on a line
parallel to the $[111]$ direction.
The non-stoichiometric QDs of the types (a) and (b) both have
centers of inversion
and are characterized by the cube symmetry group $O_h$.
The stoichiometric QD of the type (c) has no inversion center and
is characterized by the tetrahedron symmetry group $T_d$.
Note that the QDs cannot be perfectly spherical due to the
discretness and lower point symmetry
of the underlying crystal lattice. In our work the QDs
are formed
by all the atoms
within a certain distance from the center of the NC.
It is convenient to measure this distance with a dimensionless
integer number. Thus, we define the ``number of shells'' as
the number of atomic layers within the distance from the
center of the QD to its surface along the $[100]$ direction.

Contrary to covalent semiconductors like Si~\cite{Delerue2004b},
lead chalcogenides are characterized by the strongly ionic
atomic bonds making them relatively insensitive to the surface chemistry~\cite{klimeck2011}.
In particular, no surface states appear in the
fundamental band gap of non-passivated lead chalcogenide NCs.
Therefore, we have not passivated the surface atoms in our TB
modeling. In real QDs of the types (a) and (b)
(cf.~Fig.~\ref{fig:structure})
such passivation is necessary to compensate for the surface charge~\cite{Leitsmann2009}. The actual structure of the NCs should depend on the
details of the synthesis procedure and can be determined
with the help of the nuclear magnetic resonance\cite{Moreels2008} and X-ray diffraction\cite{Petkov2010} techniques.

The calculated energy levels of confined carriers for PbSe and PbS
NCs of the diameter $D \approx 5$~nm (corresponding to 9 shells)
are shown in Fig.~\ref{fig:Dos}.
For each material three panels (a--c and a$'$--c$'$) correspond to the
three possible NC
geometries illustrated in~Fig.~\ref{fig:structure}.
The band gap in both cases agrees well with the results
of Ref.~\cite{Delerue2004b}.

All the states can be divided into distinct groups characterized by
a certain parity.
For NCs with a center of inversion, each state automatically
has a certain parity. Indeed, in bulk lead chalcogenides the lowest
electron state in the conduction band
has the $L_6^-$ symmetry~\cite{Cardona2010},
{\it i.e.} it is odd with respect to the inversion symmetry operation
when the center of inversion is
chosen on the cation atom.
The uppermost electron state in the valence band has the opposite parity.
For QDs without an inversion center one can define approximate projectors
to the even and odd states. We have attributed a certain parity to the states,
for which the squared mean values of the projectors differed more
than in three times.

Energy splittings within each multiplet characterized by a certain parity
 are clearly seen in~Fig.~\ref{fig:Dos}
and can be explained by the confinement-induced
inter-valley coupling and the carrier effective mass
anisotropy.
The importance of these two effects for lead chalcogenide NCs
has been emphasized in Ref.~\cite{zunger2006}.
However, the dependence of the splittings on
the NC geometry clearly manifested in~Fig.~\ref{fig:Dos}
has never been reported.

A striking feature of Fig.~\ref{fig:Dos} is the  suppression of
the energy splittings for the type (c) NCs
lacking a center of inversion.
The splittings are quite small and cannot be distinguished within the 
energy scale of Fig~\ref{fig:Dos}. On the contrary, for QDs with an inversion center
 [panels (a), (b), (a$'$), (b$'$) of Fig.~\ref{fig:Dos}], the ground-state multiplets 
for both electrons and holes have well-pronounced structures with 
substantial splittings even for QD diameters as large as 4.9 nm. This 
observation refers to both the conduction and valence band
electron states.
The effect is more pronounced for the PbS QDs than
for the PbSe QDs, which can be related to the more
isotropic effective masses of the band extrema
in bulk PbS.

\begin{figure}[t!]
 \includegraphics[width=\linewidth]{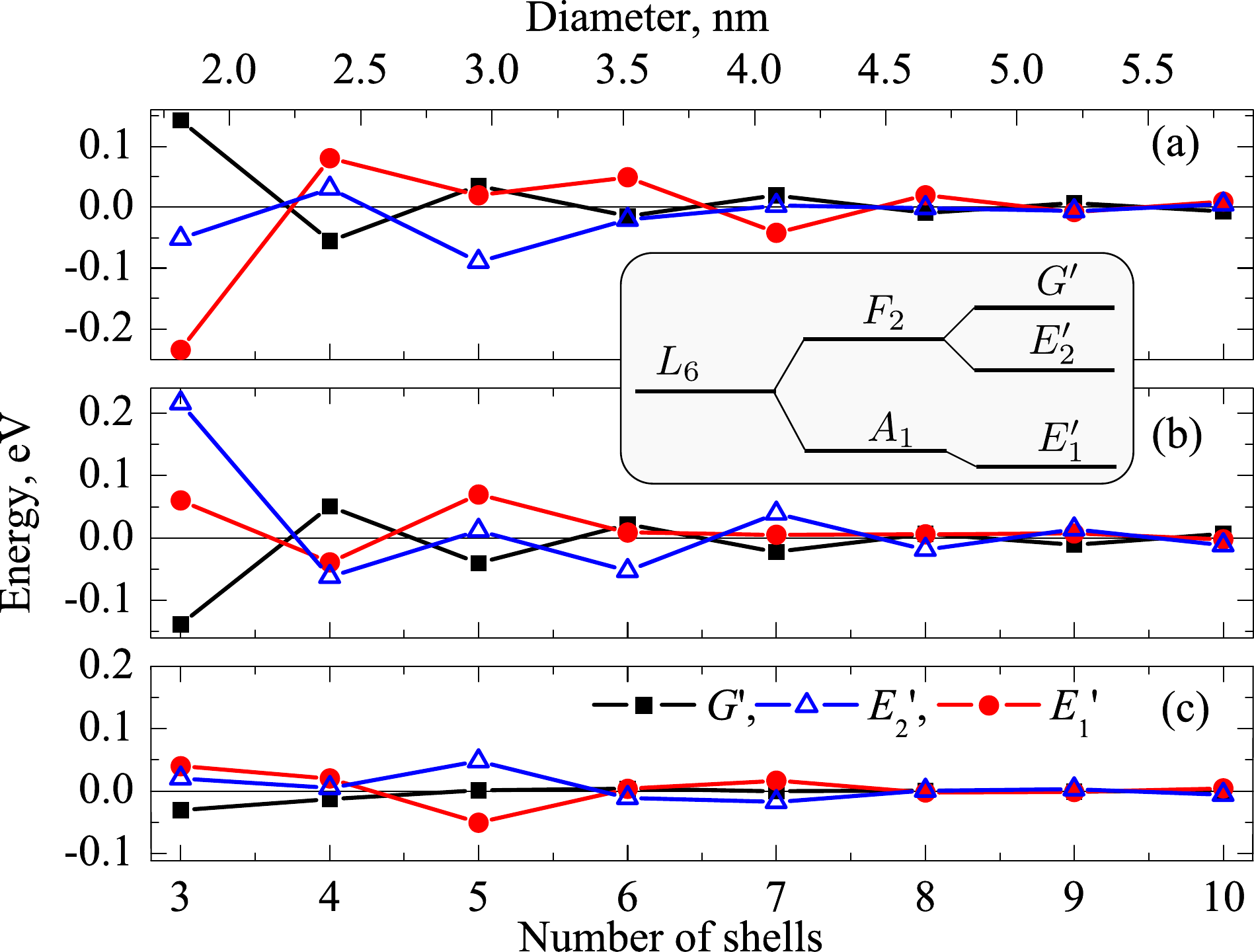}
\caption{(Color online) Energies of levels belonging 
to the ground state multiplet of the conduction band electron in PbS NCs as functions of NC diameter.
Panels (a), (b), (c)  correspond to Pb-centered NCs, S-centered NCs and NCs without inversion center, see Fig.~\ref{fig:structure}.
Squares, triangles and circles correspond to the states with the symmetry $G'$,  $E_2'$ and $E_1'$, respectively, see the level splitting scheme in the inset.
}\label{fig:ValleysE}
\end{figure}

\begin{figure}[b!]
 \includegraphics[width=\linewidth]{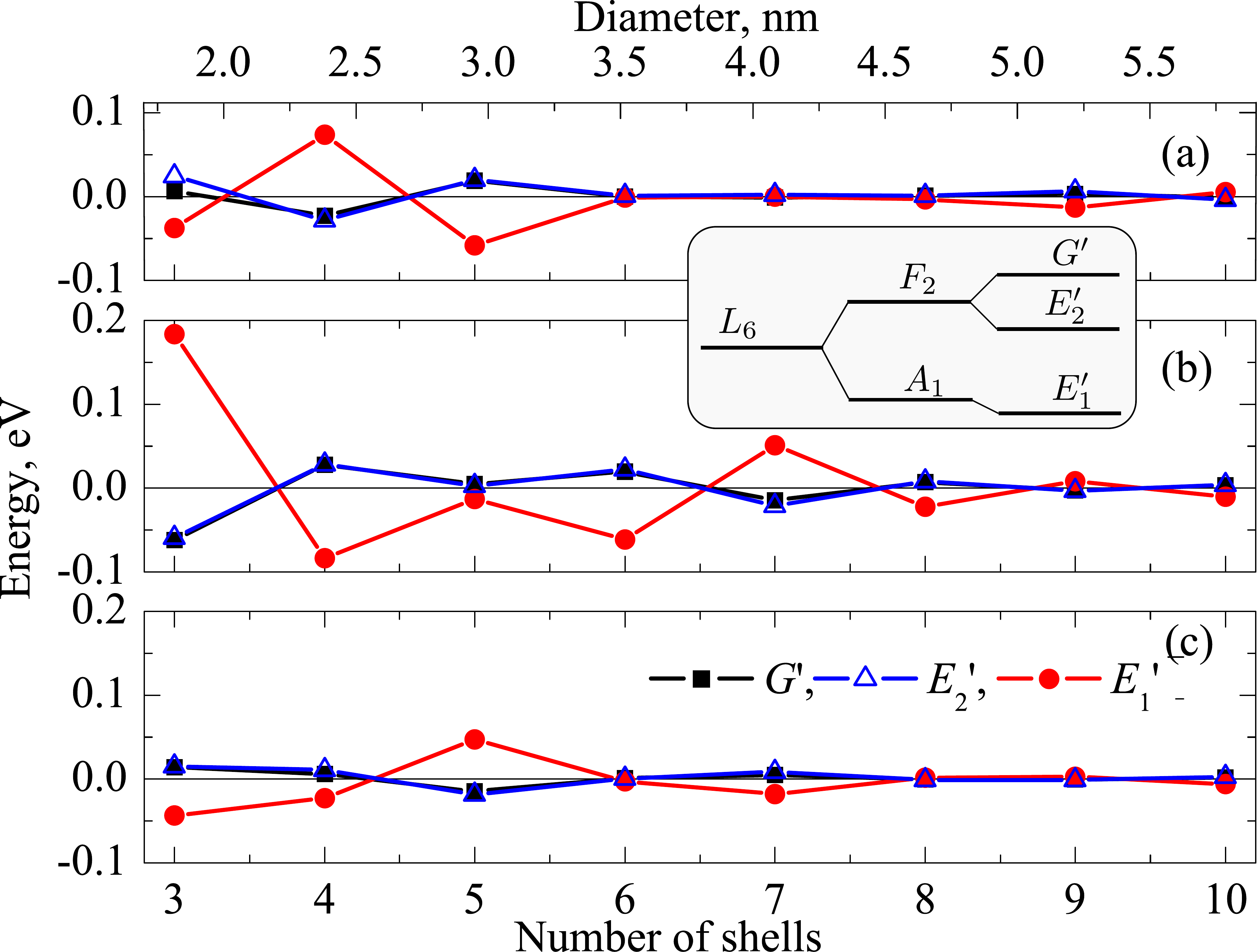}
\caption{(Color online)
Same as Fig.~\ref{fig:ValleysE}, but for the valence-band ground state.
}\label{fig:ValleysH}
\end{figure}

To elucidate this puzzling behavior, we have analyzed the
dependence of the splittings on the NC diameter.
For simplicity, we restrict our consideration by the
electron and hole ground states.
Within the effective mass approximation, the ground state
of confined carriers is fourfold degenerate with respect
to the valley index and twofold degenerate with respect
to the spin projection, {\it i.e.} the total degeneracy is eightfold.
If we neglect the spin and consider valley-orbit interaction only,
the ground state is split into a state of $A_1$ symmetry (singlet),
and a state of $F_2$ symmetry (triplet), as sketched in insets of Figs.~\ref{fig:ValleysE},\ref{fig:ValleysH}.
When the spin degree of freedom is taken into account
then both the singlet and the triplet states acquire extra
degeneracy. This degeneracy is partly lifted, as the six-fold
degenerate state corresponding to the triplet is split by the
spin-orbit interaction into a two-fold degenerate state
of $E_{2}'$ symmetry and a four-fold degenerate state
of $G'$ symmetry~\cite{BirPikus}.
As a result, the carrier ground-state level is split into
the three multiplets: the two doublets (of $E_{1}'$ and $E_{2}'$ symmetry,
respectively) and the
four-fold degenerate state of $G'$ symmetry.
As far as the symmetry with respect to inversion is not concerned, the symmetry groups $T_d$ and $O_h$ are equivalent.
Therefore, this symmetry analysis applies to all types of NC geometries presented in Fig.~\ref{fig:structure}.

Figures \ref{fig:ValleysE} and \ref{fig:ValleysH} show the energies of
the resulting conduction (valence) band
multiplets in PbS NCs as functions of the
NC diameter.
The panels (a)--(c) correspond to the three NC geometries
considered throughout the paper (see Fig.~\ref{fig:structure}).
The energies of the states are counted from the averaged value
$(E_{E_1'}+E_{E_2'}+2E_{G'})/4$.
The splittings strongly oscillate with the number of shells $N$ in a NC.
Such oscillations are typical for the valley splittings in various
semiconductor structures. Similar behavior has been reported for
SiGe/Si\cite{nestoklon1,frisen2007} and GaSb/AlAs\cite{ting1988,jancu2004}
quantum wells and Si NCs~\cite{Bulutay2007}.


Comparison of panels (a) and (b) of~Figs.~\ref{fig:ValleysE},\ref{fig:ValleysH}
on one hand with the panels (c) of~Figs.~\ref{fig:ValleysE},\ref{fig:ValleysH}
on the other hand clearly shows that the suppression of valley
splittings in NCs without a center of inversion is a general
feature persistent in a wide range of NC sizes.
Comparison of~Fig.~\ref{fig:ValleysE} with~Fig.~\ref{fig:ValleysH} enables one
to conclude that the spin-orbit interaction is much stronger for
conduction-band electrons than for valence-band holes. Indeed,
the energies of the hole states with the symmetry $G'$ and $E_2'$
in~Fig.~\ref{fig:ValleysH} are almost the same, while,
for conduction-band electrons, all the splittings
in~Fig.~\ref{fig:ValleysE}
are of the same order.

Let us explain the
anomalous suppression of the valley splittings in lead salt NCs
without an inversion center. We want to account for the inter-valley
coupling in the lowest non-vanishing approximation. To this end we
consider electronic states originating from the four inequivalent $L$ valleys
of a bulk semiconductor and neglect the ${\bm k} \cdot {\bf p}$ mixing of
conduction and valence band states~\cite{Kang1997}. Then the wave function
of the confined electron state in the $j$-th $L$-valley can be written as
$
 \langle\bm r|L_j \rangle=\e^{\i\bm k_j\bm r}u_{j}(\bm r)\Phi_j(\bm r)\:,
$
where $\Phi_j(\bm r)$ is the smooth envelope function,
$u_j(\bm r)$ is the periodic Bloch amplitude for the bulk state
in the $j$-th valley, and spin indices are omitted.
We will further assume that the bulk material has
isotropic effective masses of the band extrema in
$L$ points.
In this approximation
the envelopes $\Phi_j(\bm r)$ are invariant under rotations.
The confinement-induced inter-valley coupling
can
be described by the following matrix element:
\begin{equation}
 I_{j,k}=\langle L_{j}|H_{\rm QD}|L_k \rangle,\quad  j,k=1\ldots 4,
 \quad j\ne k \,,
\end{equation}
where $H_{\rm QD}$ is the microscopic QD Hamiltonian.
Then it follows that
the integral $I_{j,k}$ vanishes
when the QD lacks inversion symmetry,
{\it i.e.} belongs to the type (c).
To show this let us rewrite $I_{j,k}$ as
\begin{multline}\label{eq:ijk2}
I_{j,k} \approx \int\limits_{u. c.} \e^{-\i \bm k_j \bm r} \,
u_j^*(\bm r) \, H_{bulk}  \, \e^{\i \bm k_k \bm r}
\, u_k(\bm r) d \bm r \,
\\ \times \sum\limits_{{\bf R}_n} \e^{\i(\bm k_k-\bm k_j) {\bm R}_n} \,
\Phi^*_j({\bm R}_n) \, \Phi_k({\bm R}_n) \,,
\end{multline}
where the integral in the right-hand side is over a unit cell
and contains the Hamiltonian of a bulk material
while the summation runs over all the cation (or anion) sites
{\it within the QD}. It is this summation that is sensitive to the
arrangement of atoms within the QD.
For the type (c) geometry the sum is exactly zero.
This cancellation takes place independently of the radius of the
QD and is fully determined by the symmetry.
To see this one can use the following well known fact~\cite{BirPikus}.
If a given function describing
some crystalline physical system
transforms according to a
certain representation of the system's symmetry group,
then the sum of this function over the lattice sites belonging
to the system may be different
from zero if and only if the decomposition of this representation
into irreducible ones contains the identity representation.

In our case one can distinguish three linearly independent functions
$\exp[\i(\bm{k}_k-\bm{k}_j) {\bm R}_n]$ which may be chosen as shown
in the first column of Table~\ref{table2}. Table~\ref{table2}
gives the values of these exponent functions when ${\bm R}_n$ sweeps
the coordinates of the anion atoms shown in Fig.~\ref{fig:structure}(c).
These atoms may be obtained from one another by the rotations of the
type (c) QD. The last column of Table~\ref{table2} indicates that the
exponent functions transform according to the vector irreducible
representation $F_2$ of the group $T_d$. This representation is different
from the identity representation $A_1$. Therefore, for type (c) QDs
Eq.~(\ref{eq:ijk2}) is zero. Table~\ref{table3} gives the values of the
same exponent functions when ${\bm R}_n$ sweeps
the coordinates of the anion atoms shown in Fig.~\ref{fig:structure}(a).
These atoms may be obtained from one another by the rotations of the
type (a) QD. The last row of Table~\ref{table3} shows that the sum of
the exponent functions remains invariant under such rotations. More
precisely, the exponent functions
transform according to the direct sum of the two irreducible
representations $A_1^+ \oplus E^+$ of the group $O_h$.
Thus, for type (a) QDs
Eq.~(\ref{eq:ijk2}) is different from zero.

\begin{table}[ht!]
 \caption{Phase factors in Eq.~\eqref{eq:ijk2} for the
coordinates of the anion atoms in Fig.~\ref{fig:structure}(c).
}\label{table2}
\begin{tabular}{c|c|c|c|c|c}
&$\frac{a}{4}(1,1,1)$&$\frac{a}{4}(1,\bar 1,\bar 1)$&$\frac{a}{4}(\bar 1,1,\bar 1)$&$\frac{a}{4}(\bar 1,\bar 1,1)$&\mbox{}\\\hline
 $\e^{\i({\bm k}_1-{\bm k}_2){\bm R}_n}$&-1&-1&1&1&$-x$\\
 $\e^{\i({\bm k}_1-{\bm k}_3){\bm R}_n}$&-1&1&-1&1&$-y$\\
$\e^{\i({\bm k}_1-{\bm k}_4)\bm R_n}$&-1&1&1&-1&$-z$
\end{tabular}
\end{table}
\begin{table}[th!]
\caption{ Same as Table~\ref{table2} but for anion atoms in Fig.~\ref{fig:structure}(a).
}\label{table3}
\begin{tabular}{c|c|c|c|c}
&$(\pm\frac{a}{2},0,0)$
&$(0,\pm\frac{a}{2},0)$
&$(0,0,\pm\frac{a}{2})$
&\mbox{}
\\\hline
 $\e^{\i({\bm k}_1-{\bm k}_2){\bm R}_n}$&1&-1&-1&\mbox{}\\
 $\e^{\i({\bm k}_1-{\bm k}_3){\bm R}_n}$&-1&1&-1&\mbox{}\\
$\e^{\i({\bm k}_1-{\bm k}_4){\bm R}_n}$&-1&-1&1&\mbox{}\\\hline
$\sum_{j=2}^4\e^{\i({\bm k}_1-{\bm k}_j){\bm R}_n}$
 &-1&-1&-1&$A_1^+$
\end{tabular}
\end{table}

This consideration is no longer valid
if the function $\Phi_j(\bm{r})$ is anisotropic.
This is the case of real lead salts NCs, as in bulk
lead chalcogenides the longitudinal mass in the $L$ valley is
larger than the transverse one.
Consequently, in real NCs lacking inversion center the valley
splitting is not exactly zero but
determined by the degree of the effective mass anisotropy
in $L$ valleys.
This explains the fact that in PbS NCs the splitting is
smaller than in PbSe ones, cf. panels (c) and (c$'$) of Fig.~\ref{fig:Dos}.
In conclusion, we obtained a new set of $sp^3d^5s^*$ TB
parameters for the bulk PbSe and PbS semiconductor compounds
and calculated the electron and hole energy levels
in NCs made of these materials.
We demonstrated that the valley-orbit and spin-orbit splittings
of the ground electron and hole energy levels
are
very sensitive to
a particular arrangement of atoms in the NC and can be strongly
suppressed for a certain geometry, when the NC lacks a
center of inversion.

\acknowledgments Useful discussions with E.L. Ivchenko and L.E. Golub are gratefully acknowledged.
This work was supported by the Russian Foundation for Basic Research, European projects POLAPHEN and Spin-Optronics and the ``Dynasty'' Foundation-ICFPM.
The work of SVG was supported, in part, by the Research Corporation for Science Advancement under
Award No.~20081 and, in part, by the National Science Foundation under Grant No. HRD-0833178. The work of MON was partially supported by ``Triangle de la Physique''.

\end{document}